\UseRawInputEncoding
\pdfoutput=1
\documentclass[a4paper,11pt]{article}
\pdfoutput=1 

\usepackage{jcappub} 

\usepackage[T1]{fontenc} 
\usepackage{graphicx}
\usepackage{dcolumn}
\usepackage{bm}
\usepackage{natbib}
\usepackage[mathlines]{lineno}
\usepackage{amsmath}

\def\be{\begin{equation}}
\def\ee{\end{equation}}
\graphicspath{ {figure/} }

\title{\boldmath Revisiting Bounds on Primordial Black Hole as Dark Matter with X-ray Background}

\author[a]{Xiu-Hui Tan}
\author[a,b,1]{Jun-Qing Xia\note{Corresponding author.}}

\affiliation[a]{Department of Astronomy, Beijing Normal University, Beijing 100875, China}
\affiliation[b]{Institute for Frontiers in Astronomy and Astrophysics, Beijing Normal University, Beijing 100875, China}

\emailAdd{tanxh@bnu.edu.cn}
\emailAdd{xiajq@bnu.edu.cn}

\abstract{Within the mass range of $10^{16}-5\times 10^{18}$ g, primordial black holes (PBHs) persist as plausible candidates for dark matter. Our study involves a reassessment of the constraints on PBHs through a comparative analysis of the cosmic X-ray background (CXB) and the emissions arising from their Hawking evaporation. We identify previously overlooked radiation processes across the relevant energy bands, potentially refining the bounds on PBHs. These processes encompass the direct emission from Hawking radiation, in-flight annihilation, the final state of radiation, and positronium annihilation. Thorough consideration is given to all these processes and their respective emission fractions, followed by a precise calculation of the $\mathcal{D}$ factor for observations directed towards the high-latitude Galactic contribution. Furthermore, we integrate the flux originating from extragalactic sources, both of which contribute to the measured isotropic flux. Through a comparative analysis of data derived from previous CXB observations utilizing an Active Galactic Nuclei (AGN) double power-law model, we establish the most stringent constraints for PBHs, thereby excluding the possibility of PBHs constituting the entire dark matter mass within the range of $2.5\times 10^{17}-3 \times 10^{17} $g.}

\begin{document}
\maketitle
\flushbottom

\section{\label{sec: into}Introduction} 
Primordial Black Holes (PBHs) as the exclusive candidate for Dark Matter (DM) without necessitating novel physics has been extensively explored in the academic literature \cite{carr_primordial_2016}. In contrast to particle-based DM, PBHs represent theoretical entities originating from the gravitational collapse of stochastic quantum overdensities during the early stages of the Universe, displaying a more macroscopic character. Hawking radiation \cite{1975CMaPh..43..199H,10.1093/mnras/152.1.75} emitted by Primordial Black Holes can generate a detectable signal that contributes to the observed cosmic X-ray background (CXB). Currently, PBHs would entirely evaporate if their mass fell below approximately $10^{15}$ g. With increasing PBH mass, the energy spectrum tends to soften, rendering PBHs larger than $5\times 10^{18}$ g negligible contribution. Consequently, a notable knowledge gap in our understanding of PBHs within the mass range of $10^{16}$ to $5\times 10^{18}$ g, particularly concerning their possible existence amidst asteroid masses.

Previous investigations have yielded crucial constraints on the mass of PBHs and their fraction ($f_{\rm PBH}=\Omega_{\rm PBH}/\Omega_{\rm DM}$) as constituents of Dark Matter. For instance, using data from the \textit{INTEGRAL} satellite on Galactic $\gamma$-ray emission, ref.\cite{laha_integral_2020} demonstrated that PBHs with a mass of $M_{\rm PBH}=1.2\times 10^{17}$ g fall short of constituting the entirety of Dark Matter. Furthermore, in elucidating the galactic 511 keV line detected by \textit{INTEGRAL/SPI}, refs.\cite{Bambi:2008kx, DeRocco:2019fjq, Laha:2019ssq, Keith:2021guq} excluded specific parameter spaces for PBHs. Ref.\cite{cai_constraints_2021} even employed a mixing model incorporating PBHs and DM particles to further constrict the parameters of PBHs. The potential contribution of PBHs to the extant cosmic X-ray background or isotropic $\gamma$-ray background has also been scrutinized in studies by refs.\cite{Iguaz:2021irx, Korwar:2023kpy, Chen:2021ngo}, where they explored photons stemming from electron-positron pair-annihilation and refined the constraints on PBHs to approximately $\sim 2\times 10^{17}$g, underscoring the significance of evaporated positrons in shaping the cosmic X-ray background. Several other studies have forecasted constraints on PBHs through forthcoming telescopes \cite{Keith:2022sow, ray_near_2021, Tan:2022lbm}.

This study presents a reassessment of the X-ray diffuse emission originating from PBHs spanning the mass range of $10^{16}-5\times 10^{18}$ g. In contrast to prior analyses, we conduct a comprehensive examination of all potential emissions contributing to the isotropic flux. These include: 1) Photons directly emitted from Hawking radiation, encompassing primary and secondary prompt emission components; 2) Photons generated from internal bremsstrahlung radiation; 3) Positrons produced by PBHs, which interact with electrons, leading to photon generation through three distinct processes; 4) Differentiating contributions from the Galactic (Gal) anticenter and extragalactic (EG) regions. Furthermore, we meticulously assess the respective proportions of each emission mode and collate all available CXB measurements to delineate the parameter space accurately.

The paper is structured as follows: Section \ref{sec: emission} presents a detailed account of the computational specifics for each emission mode and their respective fractions, which are subsequently aggregated to determine the overall contributions. We describe the flux originating from two distinct regions, namely Gal and EG, in Section \ref{sec: region}. Our analysis and constraint outcomes are expounded upon in Section \ref{sec: analy}, by comparing with the CXB data. Finally, our findings are deliberated upon and summarized in Section \ref{sec: concl}.

\section{Emission classification}\label{sec: emission}
According to the Hawking Evaporation theory, it is anticipated that the asteroid mass of PBHs can be inferred from the X-ray and soft $\gamma$-rays emissions. The temperature of a Schwarzschild black hole can be correlated to its mass through the following expression:
\be
T_{\rm BH} = \frac{M_P^2}{8 \pi M_{\rm BH}},
\ee
where $M_P \approx 1.22\times 10^{19}~\rm{GeV}$ denotes the Planck mass. The black hole emits particle species $i$ with a species-specific greybody factor $\Gamma_i(E, M_{\rm BH})$, resembling blackbody radiation. This factor encapsulates the probability of a Hawking particle escaping the gravitational potential well of the PBHs. 
Consequently, the number density of particle $i$ can be computed using the following expression:
\be
\frac{{\rm d}N_i}{{\rm d}E{\rm d}t}=\frac{1}{2\pi}\frac{\Gamma_i(E, M_{\rm BH})}{e^{E/T_{\rm BH}}-(-1)^{2s}},
\label{equ:greybody_emit}
\ee
where $s$ represents the spin of the emitted particle and $E$ signifies its energy. A publicly available computational tool, \texttt{BlackHawk} \cite{Arbey:2019mbc, Auffinger:2020ztk}, is capable of producing numerical datasets within the pertinent energy spectrum emitted by PBHs. The contributions of different particle species $i$ to the isotropic X-ray flux may exhibit variations. 

In our computations, we account for all potential photon emissions. It is imperative to integrate over the relevant mass range when considering a specific mass distribution of PBHs. Although we assume a monochromatic mass distribution for PBHs, other distributions such as log-normal or power-law models are more realistic. Previous study \cite{Carr:2017jsz} has demonstrated that an extended mass function could impose stricter constraints compared to a monochromatic distribution. However, due to the presence of several other free parameters in our analysis, distinguishing between these models is not straightforward. Consequently, this study focuses solely on the monochromatic mass distribution to derive the most conservative PBH constraint outcomes.

\subsection{Photons from direct radiation}
We designate the photons emitted directly from Hawking evaporation as prompt radiation, encompassing both primary and secondary emissions. Primary photons are defined by Equ.(\ref{equ:greybody_emit}), with $s=1$. The secondary component emerges from the decay of hadrons originating from the fragmentation of primary quarks and gluons \cite{Arbey:2019mbc}, as well as from the decay of gauge bosons. Furthermore, hadrons originating from PBHs also undergo decay or radiation processes. Hence, the direct number density originating from PBHs can be expressed as:
\be
\frac{{\rm d}N^{\rm dir}_{\gamma}}{{\rm d}E{\rm d}t}=\frac{{\rm d}N^{\rm prompt}_{\gamma}}{{\rm d}E{\rm d}t}+\frac{{\rm d}N^{\rm FSR}_{\gamma}}{{\rm d}E{\rm d}t} + \frac{{\rm d}N^{\rm dec}_{\gamma}}{{\rm d}E{\rm d}t},
\label{equ: direct-emit}
\ee
where the latter two terms are adopted from ref.\cite{Coogan:2020tuf}. The Final State Radiation (FSR) denotes radiation that remains independent of the astrophysical setting. The combination of FSR and virtual internal bremsstrahlung forms what is known as the full internal bremsstrahlung (IB) radiation. Nevertheless, we found some discrepancies in the studies conducted by refs.\cite{Bringmann:2012vr, Siegert:2021upf, Keith:2022sow}, while the original formulation of IB radiation was presented by refs.\cite{Beacom:2005qv, beacom2005gamma}. Given these circumstances, we are inclined to favor the standard particle process outlined in ref.\cite{Coogan:2020tuf}, which is detailed as follows:
\be
\begin{split}
    \frac{{\rm d}N^{\rm FSR}_{\gamma}}{{\rm d}E{\rm d}t} + \frac{{\rm d}N^{\rm dec}_{\gamma}}{{\rm d}E{\rm d}t} &=\sum_{i=e^{\pm},\mu^{\pm},\pi^{\pm}} \int {\rm d}E_i \frac{{\rm d}N_i^{\rm pri}}{{\rm d}E{\rm d}t}\frac{{\rm d}N_i^{\rm FSR}}{{\rm d}E}
     + \sum_{i=e^{\pm},\pi^{0},\pi^{\pm}} \int {\rm d}E_i \frac{{\rm d}N_i^{\rm pri}}{{\rm d}E{\rm d}t}\frac{{\rm d}N_i^{\rm dec}}{{\rm d}E},
\end{split}
\label{equ: FSR_dec}
\ee

By means of computational analysis, it has been ascertained that the significant contribution to the X-ray band predominantly originates from $e^-e^+$ pairs generated by FSR, rather than decay. The comprehensive expression for FSR is as follows:
\be
\frac{{\rm d}N_i^{\rm FSR}}{{\rm d}E_\gamma}=
\frac{\alpha}{\pi Q_i}P_{i\to i \gamma}(x)\left[\log(\frac{1-x}{\mu_i^2})-1\right],
\ee
and
\be
P_{i \to i \gamma}(x) =
\left\{
\begin{aligned}
& \frac{2(1-x)}{x}, &i=\pi^{\pm} \\
& \frac{1+(1-x)^2}{x}, &i=\mu^{\pm}, e^{\pm} 
\end{aligned}
\right.
\ee
where the fine structure constant $\alpha=1/137.037$, $x=2E_\gamma/Q_i$, and $\mu_i=m_i/Q_i$, where $Q_i = 2E_i$represents the energy scale for FSR. The splitting function $P_{i\to i \gamma} (x)$ is utilized to distinguish between bosons and fermions. Consequently, Equ.(\ref{equ: FSR_dec}) has been streamlined to represent the results of $e^-e^+$ pairs in the following form:
\be
\begin{split}
    \frac{{\rm d}N_{\gamma}^{\rm FSR}}{{\rm d}E{\rm d}t} &= 
    \frac{\alpha}{2\pi}\int {\rm d}E_e \frac{{\rm d} N_e}{{\rm d} E_e {\rm d} t} \left(\frac{2}{E_e}+\frac{E_\gamma}{E_e^2}-\frac{2}{E_e}\right)
    \times \left[{\rm ln} \left(\frac{4E_e(E_e-E_\gamma)}{m_e^2} \right) \right].
\end{split}
\label{equ: FSR}
\ee

\subsection{In-flight Annihilation}
The electrons and positrons discharged from PBHs would be initially relativistic \cite{Korwar:2023kpy}. They decelerate through processes involving Compton scattering and ionization losses, with the cross-section increasing until the positrons transition to a non-relativistic state. The $\gamma$-rays spectrum resulting from the In-flight Annihilation (IA) of positrons quantifies the overall flux of annihilating $e^+$s particles before they reach a state of stopping or thermalization, as detailed in ref.\cite{beacom2005gamma}. This spectrum is expressed as follows:
	\begin{equation}
        \begin{split}
		\frac{d N_\gamma^{\mathrm{IA}}}{d E_\gamma {\rm d}t} & =\frac{\pi \alpha^2 n_H}{m_e} \int_{m_e}^{\infty} d E_{e^{+}} \frac{d N_{e^{+}}}{d E_{e^{+}}{\rm d}t} \times \int_{E_{\min }}^{E_{e^{+}}} \frac{d E}{d E / d x} \frac{P_{E_{e^{+}} \rightarrow E}}{(E^2-m_e^2)} \\
        &\times\left(-2-\frac{(E+m_e)(m_e^2(E+m_e)+E_\gamma^2(E+3 m_e)-E_\gamma(E+m_e)(E+3 m_e))}{E_\gamma^2(E-E_\gamma+m_e)^2}\right), 
        \end{split}
	\end{equation}
where the symbol $n_{\rm H}$ represents the number density of neutral hydrogen atoms. The positron spectrum emanating from PBHs, denoted as ${\rm d} N_{e^{\pm}}/({\rm d}E_{e^{\pm}} {\rm d}t)$, can be computed by employing Equ.(\ref{equ:greybody_emit}) with a parameter value of $s=1/2$. 
The rate of energy loss for positrons due to ionization interactions in the presence of neutral hydrogen, denoted as ${\rm d}E/{\rm d}x$, can be evaluated utilizing the standard Bethe-Bloch formula. 

As positrons gradually deplete a substantial portion of their energy and near their rest mass, indicated by $E_{e^+} \simeq m_e$, the probability of survival for positrons transitioning from energy $E_{e^+}$ to rest mass $m_e$ can be estimated as $P=P_{E_{e^+}\to m_e}$. The expression for this probability is provided as:
\be
P_{E_{e^+}\to E} = \exp\left(-n_H\int_E^{E_{e^+}} {\rm d}E' \frac{\sigma_{\rm ann}(E')}{|{\rm d}E'/{\rm d}x|} \right),
\label{equ: Prop_IfA}
\ee
where $\sigma_{\rm ann}$ denotes the cross-section for the annihilation of a positron with an electron at rest. The probability $P_{E_{e^+}\to E}$ for positrons with energies below a few MeV consistently falls within the range of approximately 0.95 to 1, indicating that only a minor fraction of these positrons undergo annihilation before transitioning to a non-relativistic state. For instance, at injection energies of 10 MeV, the deviation of $P$ from unity is roughly 11\% \cite{Beacom:2005qv}. Additionally, ref.\cite{Guessoum:2005cb} have computed the proportion of positrons forming positronium in flight within various neutral mediums, revealing that in-flight annihilation contributes minimally to X-ray emission. Nonetheless, we include this component for the sake of comprehensiveness.
\begin{figure}[htp]
    \centering
    \includegraphics[width=9cm]{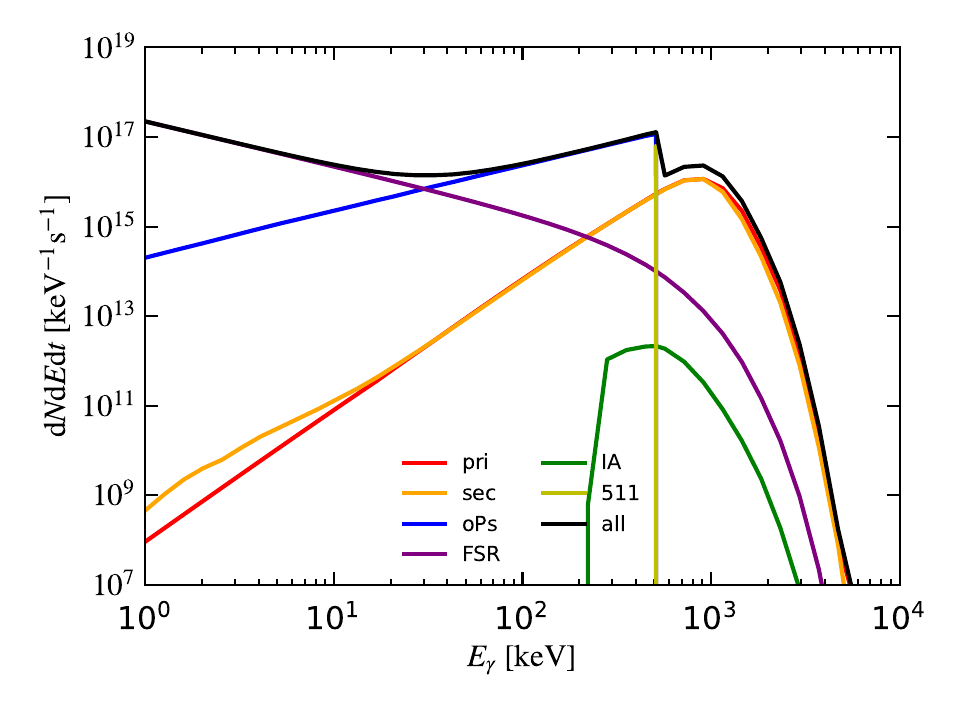}
    \caption{The number density per energy per second from a PBH with $M_{\rm PBH} = 7 \times 10^{16}$ g and $f_{\rm PBH} = 1$, encompassing all radiation sources detailed in Section \ref{sec: emission}: direct evaporation (primary and secondary emissions depicted by red and orange lines, respectively), subsequent annihilation of positron-electron pairs (blue line), final-state radiation (purple line), in-flight annihilation (green line), along with the $511~{\rm keV}$ line emission (yellow line).} 
    \label{fig: PBHs_dndedt}
\end{figure}

\subsection{\label{sec: Ps}Photons from Positronium}
The interaction between a positron and an electron results in the conversion of rest mass energy into two or more photons, with a total energy of $1022$ keV. In the scenario where a pair of electrons and positrons are at rest, their direct annihilation gives rise to two photons, each possessing an energy of $511$ keV. The ground state of positronium (Ps) exhibits two discernible spin configurations, contingent upon the relative orientations of the electron and positron spins. 

In specific terms, one of these configurations, referred to as para-positronium (pPs), materializes $1/4$ of the time, showcasing antiparallel spins and a total spin of $S=0$, identified as $^1S_0$. This state emits two photons moving in opposite directions, each carrying an energy of $511$ keV, akin to the outcome of direct electron-positron annihilation. The second configuration is ortho-positronium (oPs), characterized by parallel spins and a total spin of $S=1$, denoted as $^3S_1$, and occurs $3/4$ of the time. The oPs state necessitates a final state involving more than two photons. The conservation of momentum allocates the total energy of $1022$ keV among three photons, leading to a continuous spectrum of energy levels extending up to $511$ keV \cite{Prantzos:2010wi}.

The one-loop Quantum Electrodynamics (QED) correction to the spectrum in oPs decay has been computed numerically in ref.\cite{Manohar:2003xv, Ore:1949te}, where the Ore-Powell result is provided:
\be
\begin{split}
    \frac{1}{\Gamma_0}\frac{{\rm d}\Gamma}{{\rm d} x} & = \frac{2}{\pi^2-9} \left[\frac{2-x}{x}+\frac{(1-x)x}{(2-x)^2} \right. 
 \left. -\frac{2(1-x)^2 {\rm log}(1-x)}{(2-x)^3} +\frac{2(1-x){\rm log}(1-x)}{x^2} \right].
\end{split}
\ee
In order to differentiate it from parallel positronium, we isolate the expression $\frac{{\rm d}N^{\rm oPs}_\gamma}{{\rm d}E{\rm d}t}$, which signifies the continuous flux resulting from the three-photon decay of ortho-positronium. Subsequently, we obtain:
\be
\frac{d N_\gamma^{\mathrm{oPs}}}{d E_\gamma {\rm d}t} =  \frac{1}{\Gamma_0}\frac{{\rm d}\Gamma}{{\rm d} x} \int_{m_e}^\infty {\rm d}E\frac{{\rm d}N_e}{{\rm d}E{\rm d}t},
\ee
where $x = E/m_e$, and ${\rm d}N_e/{\rm d}E{\rm d}t$ denotes the positron spectrum originating from PBHs.

The $511$ keV line flux is from direct annihilation with $e^-$s and intermediate formation of positronium, and they produce two quasi-monochromatic photons at its mass, which can be written as
\be
\frac{d N_\gamma^{\mathrm{511}}}{d E_\gamma {\rm d}t} = \frac{d N_e}{d E_e {\rm d}t}\delta(E_e - m_e),
\ee
where $\delta$ is the Dirac-delta function.

\subsection{Fraction of each emission} 
To account for the prompt $\gamma$-ray emission from Hawking Evaporation and other processes involving electron and positron by-products, it is essential to delineate the fraction of each emission. The probability, symbolized as $P$ as previously indicated in Equ.(\ref{equ: Prop_IfA}), denotes the chance of survival in a non-relativistic state for a positron. Through the utilization of $P$ and the probability of positronium formation $f_{\rm Ps}$, we can determine the fraction linked to the aforementioned contribution.

According to ref.\cite{Beacom:2005qv}, the flux of internal bremsstrahlung $\gamma$-rays is directly related to the rate of positron production, which can be inferred from the intensity of $511$ keV. Drawing on this information and the elucidation in Section \ref{sec: Ps}, where $f_{\rm Ps}$ represents the count of $e^+$s forming positronium before annihilation, we can derive the following fractions for each emission:
\be 
\begin{split}
     f_{\rm IA} &= 1-P, \\
     f_{\rm oPs} &= 3\times \frac{3}{4}P f_{\rm Ps}, \\
     f_{\rm IB/FSR} &= \frac{1}{2} \left(1-\frac{3}{4} f_{\rm Ps}\right)^{-1}, \\
     f_{511} &= 2 \times \left[P(1-f_{\rm Ps})+ \frac{1}{4}P f_{\rm Ps}\right]. \\
\end{split}
\label{equ: fraction}
\ee
The emitted photons are multiplied by the number of oPs and $511$ keV, thus yielding $f_{\rm IA}+f_{\rm oPs}/3+f_{511}/2=1$, all originating from electrons or positrons. We introduce the total annihilation flux, which encompasses these three elements:
\be
\frac{{\rm d}N^{\rm Ann}_\gamma}{{\rm d}E {\rm d}t}  = \frac{{\rm d}N^{511}_\gamma}{{\rm d}E {\rm d}t} + \frac{{\rm d}N^{\rm IA}_\gamma}{{\rm d}E{\rm d}t} + \frac{{\rm d}N^{\rm oPs}_\gamma}{{\rm d}E{\rm d}t}.
\label{equ: ann_emit}
\ee

Consequently, the final contribution can be computed in the subsequent manner:
\be
\begin{split}
    \frac{{\rm d}N^{\rm tot}_\gamma}{{\rm d}E{\rm d}t}
    &=\frac{{\rm d}N^{\rm dir}_\gamma}{{\rm d}E{\rm d}t} + \frac{{\rm d}N^{\rm Ann}_\gamma}{{\rm d}E{\rm d}t}\\
    &=f_{\rm prompt}\frac{{\rm d}N^{\rm prompt}_\gamma}{{\rm d}E{\rm d}t}+ f_{\rm FSR }\frac{{\rm d}N^{\rm FSR}_\gamma}{{\rm d}E{\rm d}t} 
    + f_{511}\frac{{\rm d}N^{511}_\gamma}{{\rm d}E{\rm d}t} + f_{\rm IA}\frac{{\rm d}N^{\rm IA}_\gamma}{{\rm d}E{\rm d}t} + f_{\rm oPs}\frac{{\rm d}N^{\rm oPs}_\gamma}{{\rm d}E{\rm d}t}.
\end{split}
\label{equ: total_emit}
\ee
Here, we set $f_{\rm prompt}=1$ for aesthetic reasons. We illustrate an instance of continuous and line spectra, denoted as ${\rm d}N/{\rm d}E{\rm d}t$, for $M_{\rm PBH} = 7 \times 10^{16}$ g and $f_{\rm PBH}=1$ for PBH in Fig.\ref{fig: PBHs_dndedt}.

\begin{figure}[htp]
    \centering
    \includegraphics[width=9cm]{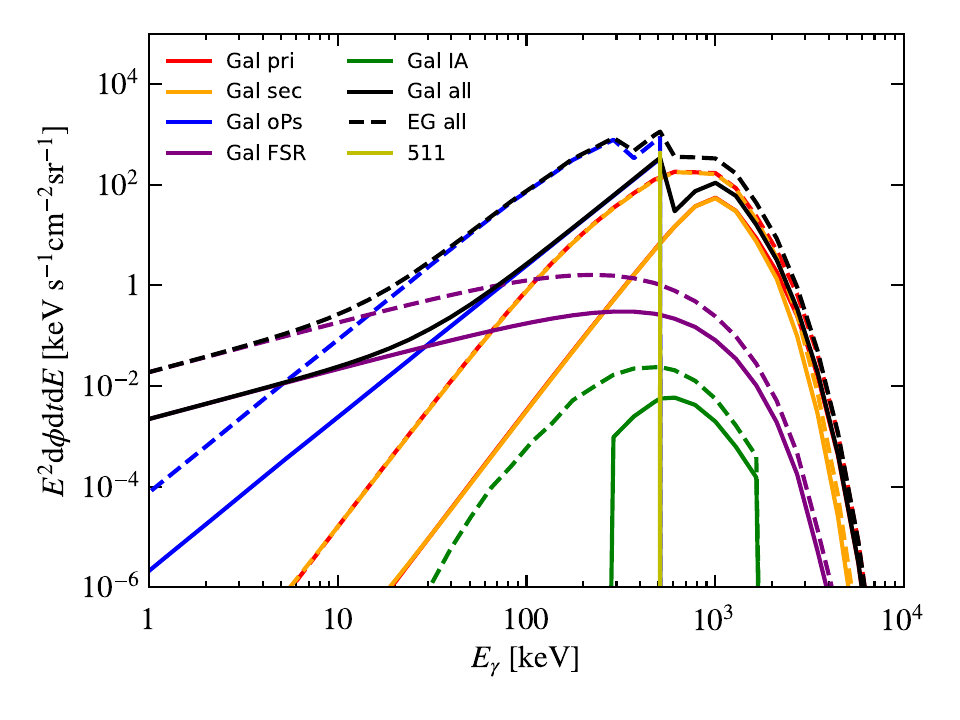}
    \caption{The expected $\gamma$-ray continuum spectra encompass various contributions to the entire isotropic differential flux, considering $M_{\rm PBH} = 7 \times 10^{16} $ g and $f_{\rm PBH}=1$. Each emission component from Gal is depicted, while only the total flux is illustrated for EG part. The legend mirrors that of Fig.\ref{fig: PBHs_dndedt}. The solid black line and dashed line signify the represent of all fluxes from Gal and EG, respectively.} 
    \label{fig: PBHs_flux}
\end{figure}

\section{\label{sec: region}Flux in two regions}
In this section, we provide an extensive overview of the complete theoretical diffuse emissions originating from PBHs, comprising two distinct components: Gal and EG,
\be
    \frac{\mathrm{d}^2\phi^{\rm PBH}}{{\rm d}t{\rm d}E{\rm d}\Omega}=\frac{\mathrm{d}^2\phi^{\rm Gal}}{{\rm d}t{\rm d}E{\rm d}\Omega}+\frac{\mathrm{d}^2\phi^{\rm EG}}{{\rm d}t{\rm d}E{\rm d}\Omega}.
\ee

Regarding the contribution from Galactic direction, we have employed a conventional Navarro-Frenk-White (NFW) profile \cite{Navarro:1995iw}. This profile is a customary benchmark selection driven by N-body simulations to characterize the galactic dark matter distribution:
\be
\begin{split}
&\rho_{\rm NFW}(r)=\rho_s \frac{r_s}{r} \left(1+\frac{r}{r_s} \right)^{-2}~{\rm with }\\
&r_s=9.98~{\rm kpc},~\rho_s=2.2\times 10^{-24}~{\rm g~cm^{-3}},\\
\end{split}
\ee
where $\rho_\odot=0.3 ~{\rm GeV~cm^{-3}}$ represents the local dark matter density, $r_\odot=8.5 ~{\rm kpc}$ denotes the distance from the solar position to the Galactic center, and $r_s$ signifies the scale radius of the dark matter halo, where $r$ signifies the distance from the Galactic center. The factor $\mathcal{D}(b,\ell)$ integrates over the line-of-sight (l.o.s), where the relationship between $s$ and $r$ is contingent on $b$ and $\ell$: $r(s, b, \ell)=\sqrt{s^2+r_{\odot}^2-2sr_\odot \cos{b}\cos{\ell} }$. The differential angular window is expressed as ${\rm d}\Omega={\rm d}(\sin b){\rm d}\ell$.

More precisely, the contributions from direct emission in Equ.(\ref{equ: direct-emit}) differ slightly from those in Equ.(\ref{equ: ann_emit}).
\be
    \frac{\mathrm{d}^2\phi^{\rm Gal}_{\gamma}}{{\rm d}t{\rm d}E{\rm d}\Omega}=\frac{f_{\rm PBH}}{4\pi M_{\rm PBH}} \left(\frac{{\rm d}^2N^{\rm dir}_{\gamma}}{{\rm d}E{\rm d}t}\mathcal{D}(b,\ell) + \frac{{\rm d}^2N^{\rm Ann}_{\gamma}}{{\rm d}E{\rm d}t}\mathcal{G}(b,\ell)\right),
\ee
Here, $\frac{{\rm d}^2N^{\rm dir}_{\gamma}}{{\rm d}E{\rm d}t}$ and $\frac{{\rm d}^2N^{\rm Ann}_{\gamma}}{{\rm d}E{\rm d}t}$ are derived from Equations (\ref{equ: direct-emit}) and (\ref{equ: ann_emit}), respectively. The $\mathcal{G}$ factor shares similarities with the $\mathcal{D}$ factor, with the primary distinction lying in its incorporation of the positron density within the Galaxy at the time of annihilation. Following the methodology outlined in ref.\cite{Iguaz:2021irx}, we made the conservative choice of setting $\mathcal{G} = \mathcal{D}$, thus defining $\mathcal{D}$ as:
\be
\mathcal{D}(b,\ell)=\int_{\rm l.o.s}{\rm d}s \rho_{\rm NFW}(s, r, \ell).
\ee
For the direction to the Gal, we determine an optical depth integral criterion stipulating that $l=180^\circ$ and $b=0^\circ$, and integral in the range of $|15^\circ|$ at the galactic anti-center, which conducives to isotropic X-ray flux.

When contemplating flux from extragalactic sources, the differentiation between direct and annihilation emissions becomes more evident, which can be expressed as:
\be
\begin{split}
    \frac{\mathrm{d}^2\phi^{\rm EG}_{\gamma}}{{\rm d}t{\rm d}E{\rm d}\Omega} & = \frac{f_{\rm PBH}\rho_{\rm c} \Omega_{\rm DM}}{4\pi M_{\rm PBH}} \int_0^{z_{\rm max}} \frac{{\rm d}z}{H(z)} 
    \times \left[\left(\frac{{\rm d}^2N_\gamma^{\rm dir}}{{\rm d}E{\rm d}t} 
    + \mathcal{F}(z)\frac{{\rm d}^2N_\gamma^{\rm Ann}}{{\rm d}E{\rm d}t}\right) e^{-\tau_\gamma(z, E(1+z))}\right],
\end{split}
\label{equ: flux_EG}
\ee
where $H(z)$ represents the Hubble constant with $H_0=67.66~{\rm km~s^{-1}Mpc^{-1}}$. We have chosen $\Omega_{\rm DM}=0.2621$ and $\rho_{\rm c}=277.5366~{\rm M_{\odot}h^2kpc^{-3}}$. The redshift integral of the extragalactic emission spans from the present time to the Cosmic Microwave Background (CMB) epoch, where $z_{\rm max}\sim \mathcal{O}(500)$. Generally, it is necessary to account for photon attenuation during propagation from the sources, where $\tau_{\gamma}\left(z, E \right)$ denotes the optical depth. We have adopted the functional form detailed in ref.\cite{Franceschini:2008tp}.

We noticed that we have introduced a suppression factor $\mathcal{F}(z)$ in Equ.(\ref{equ: flux_EG}), which considers the portion of dark matter residing within dark matter halos that contain galaxies. This factor not only confines positrons but also ensures an adequate electron density for effective positron annihilation. The adoption of this suppression factor stems from ref.\cite{Korwar:2023kpy}, who have highlighted the overlooked nat.of this formalism in various prior studies. Despite yielding a factor of $\mathcal{F}<1$, which diminishes contributions from the extragalactic component, it solely relies on dark matter halo characteristics.

A dark matter halo, characterized by a mass threshold of $M_{\rm min}(z)$, can gravitationally collapse, evolving into a significant gravitational potential well. This gravitational well plays a crucial role in attracting baryonic matter, facilitating the formation of galactic disks, and confining and annihilating DM positrons. However, cosmological simulations of galaxy formation indicate a minimum value around $\sim 10^{7}-10^{11} ~{h^{-1~M_\odot}}$. We have adopted the redshift-dependent evolution of this minimal mass from Fig.3 of ref.\cite{Rasera:2005sa}. For computation, the factor can be expressed as:
\be
\mathcal{F}(z)=\frac{1}{\rho_m(z)}\int_{M_{\rm min}}^\infty {\rm d}M M \frac{{\rm d}n(M,z)}{{\rm d}M},
\ee
where $\rho_m = \int_0^\infty {\rm d}M M \frac{{\rm d}n(M,z)}{{\rm d}M}$ for normalization, and the mass function of DM halos is based on the model by ref.\cite{Sheth:1999mn} commonly used. We showcase the flux contributions from extragalactic and Galactic sources for all components at $M_{\rm PBH} = 7 \times 10^{16}$ g in Fig.\ref{fig: PBHs_flux}, with the same legend as in Fig.\ref{fig: PBHs_dndedt}.

\begin{figure}[htp]
    \centering
    \includegraphics[width=9cm]{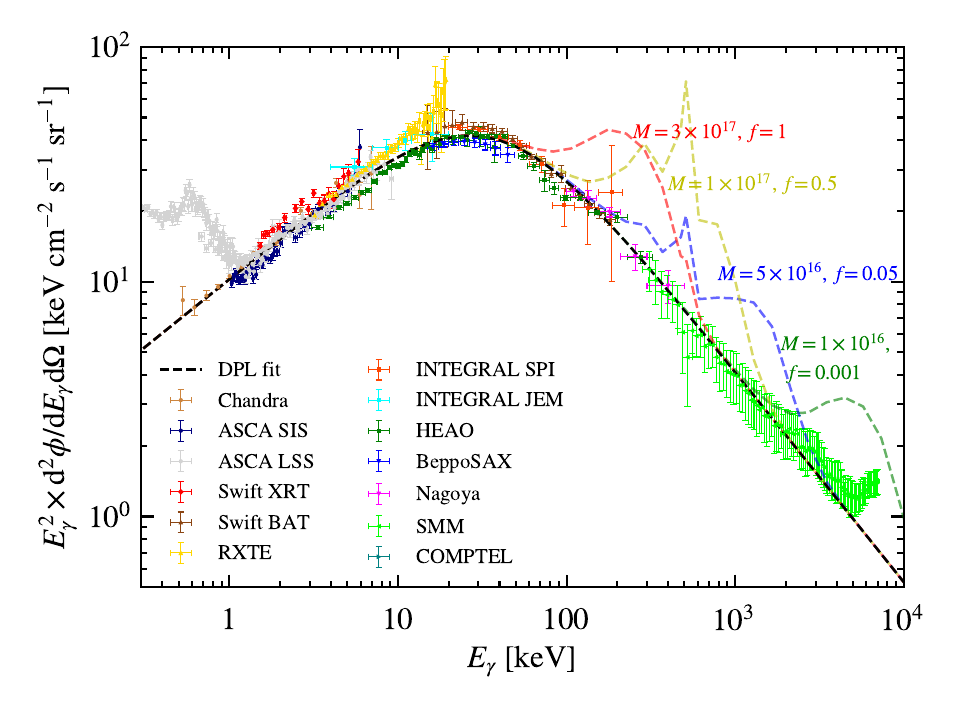}
    \caption{\label{fig: CXB}Current Cosmic X-ray background for different experiments in details, ASCA \cite{Kushino:2002vk, Gendreau95}, Integral \cite{Churazov:2006bk}, Swift \cite{Ajello:2008xb}, HEAO \cite{Gruber:1999yr}, RXTE \cite{Georgantopoulos91}, BeppoSAX \cite{Vecchi:1999ph}, SMM \cite{Watanabe97}, Nagoya \cite{fukada_rocket_1975}, Comptel \cite{Weidenspointner:2000aq}. The double power-law fit described in Equ.(\ref{equ: AGN}) is represented by the black dashed line. Additionally, corrections to the double power-law fit line are illustrated for four hypothetical monochromatic PBH spectra with varying masses in grams and cosmological abundances.} 
\end{figure}

\section{\label{sec: analy}Analysis and Results}
Many detectors are currently focused on measuring X-rays within the keV to MeV energy range. We have gathered relevant X-ray flux data ranging from $0.5$ to $400~{\rm keV}$ from sources such as ASCA \cite{Kushino:2002vk, Gendreau95}, Integral \cite{Churazov:2006bk}, Swift \cite{Ajello:2008xb}, HEAO \cite{Gruber:1999yr}, RXTE \cite{Georgantopoulos91}, BeppoSAX \cite{Vecchi:1999ph}, SMM \cite{Watanabe97}, Nagoya \cite{fukada_rocket_1975}, and Comptel \cite{Weidenspointner:2000aq}. These data encompass the CXB as depicted in Fig.\ref{fig: CXB}. Additionally, in this figure, we present four additional fluxes originating from PBHs with different masses and cosmological abundances $f_{\rm PBH}$ to adjust the double power-law (DPL) fit model.

Specifically, the CXB is commonly associated with active galactic nuclei (AGN), such as the population synthesis model proposed by ref.\cite{Ueda:2014tma}. However, it deviates notably from observed data beyond $200$ keV. Ref.\cite{Ballesteros:2019exr} introduced a proxy double power-law (DPL) fit for combined AGN and blazar emissions, providing a more consistent phenomenological form, as follows:
\be
\frac{\phi^{\rm AGN}}{{\rm d}E {\rm d}t}=\frac{A}{(E/E_b)^{n_1} + (E/E_b)^{n_2}},
\label{equ: AGN}
\ee
where the best fit parameters are determined as follows: $E_b = 35.6966$ keV, $A = 0.0642 ~{\rm keV^{-1}s^{-1}cm^{-2}sr^{-1}}$, $n_1 = 1.4199$, and $n_2 = 2.8956$. 
This fitting form is derived from models observed in the SMM, Nagoya, and HEAO experiments. It is apparent that the DPL form is not able to accurately represent data from COMPTEL and ASCA LSS as shown in Fig.\ref{fig: CXB}. Hence, we have opted to exclude them from our subsequent analysis.
\begin{figure}[htp]
    \centering
    \includegraphics[width=9cm]{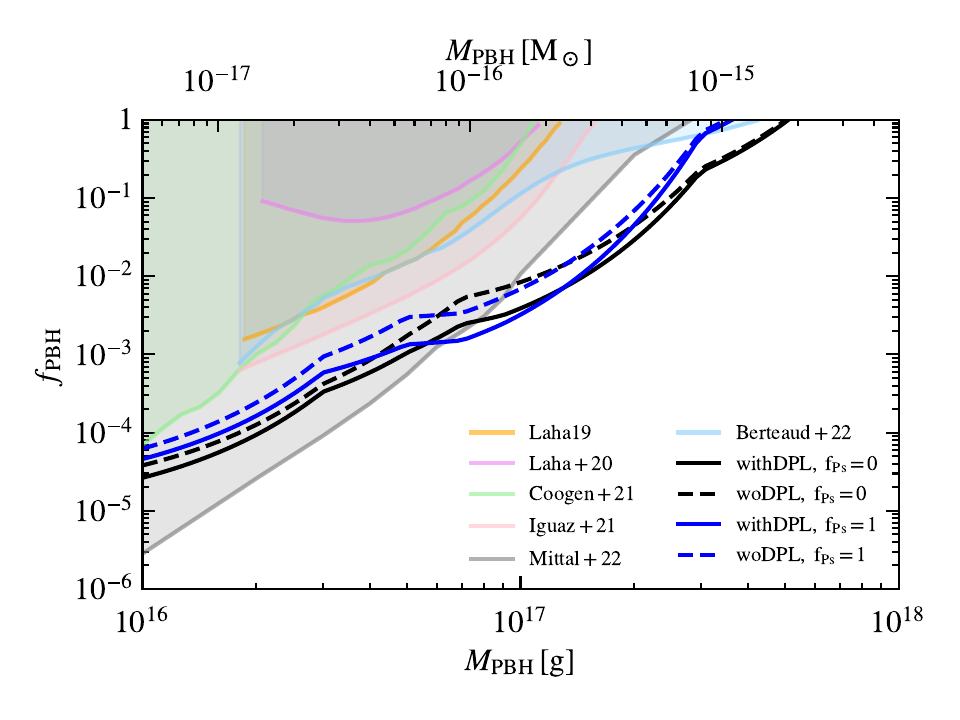}
    \caption{\label{fig: bounds_fPs}Bounds derived on the PBH abundance with respect to the PBH mass, assuming a monochromatic mass distribution. Four cases have been considered, where $f_{\rm Ps}$ takes values of 1 or 0 and includes or excludes the DPL background. It is noted that a spin value of $a=0$ has been fixed for all these cases.}
\end{figure}

\subsection{Considering of Background models}
Firstly, we consider only the PBHs model, providing a conservative estimation. Assuming that the data follows a normal distribution, the $\chi^2$ value is calculated as follows:
\be
    \chi^2=\sum\left[ \frac{\phi^{\rm{data}}_{\gamma} - \phi^{\rm PBH}_{\gamma}(\theta)}{\sigma^{\rm{data}}} \right]^2,
\ee
where $\theta = \{f_{\rm PBH},m_{\rm PBH}\}$ represents the PBH parameters, $\phi_{\gamma}$ denotes the expected integral $\gamma$-flux model from PBHs, and $\phi_{\gamma}^{\rm data}$ represents the experimental measurements with corresponding uncertainties $\sigma_{\rm data}$. The $\chi^2$ value is calculated by summing over all energy bins for the discrepancy between the data and the model.
We determined the 95\% confidence level (C.L.) bound on $f_{\rm PBH}$ for each PBH mass within the specified range. The permissible $f_{\rm PBH}$ values associated with the flux originating from PBHs should not exceed any errors of the measurement data shown in Fig.\ref{fig: CXB}, which translates to $\chi^2 - \chi^2_{\rm min} \leq \chi^2_{0.05}(N-1) \simeq 3.84$, where $N-1=1$ represents the number of degrees of freedom.

\begin{figure}[htp]
    \centering
    \includegraphics[width=9cm]{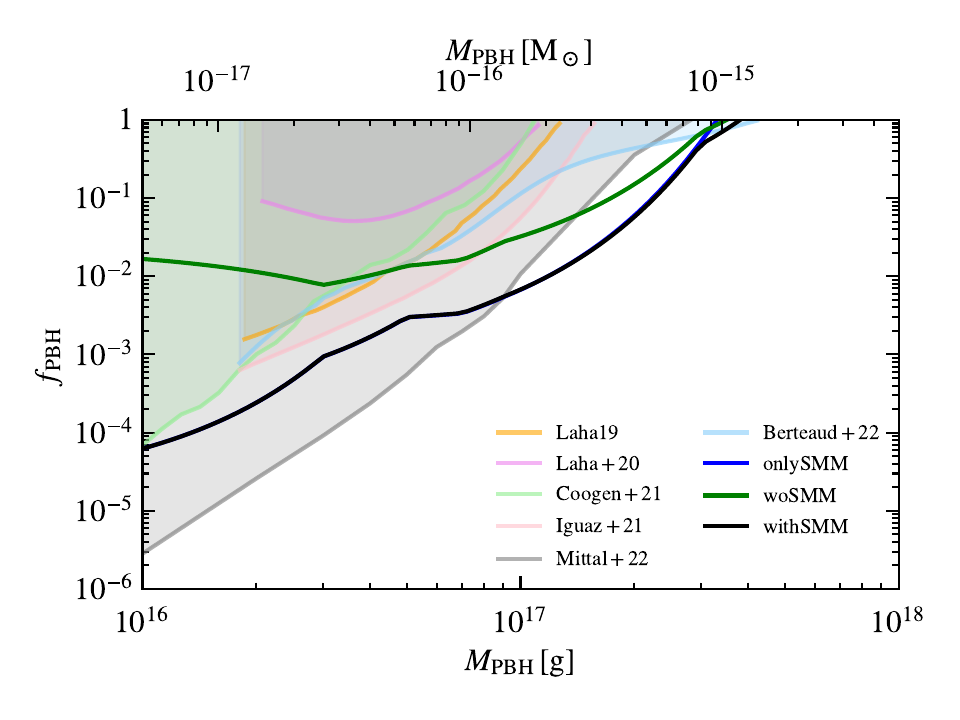}
    \caption{Bounds derived from three scenarios: only based on SMM data (onlySMM, indicated by the blue line), considering all data including SMM (withSMM, represented by the green line), and excluding SMM data (woSMM, depicted by the black line). In all three cases, the spin parameter is set at $a=0$, and $f_{\rm Ps}=1$ without the DPL model.} 
    \label{fig: bounds_woSMM}
\end{figure}

Secondly, we aim to establish a more realistic estimator for deriving the current constraint on PBHs by incorporating the contributions from the DPL fitting term $\phi^{\rm AGN}_{\gamma}$. The $\chi^2$ expression is formulated as follows:
\be
    \chi^2=\sum\left[ \frac{\phi^{\rm{data}}_{\gamma} - \phi^{\rm PBH}_{\gamma}(\theta)- \phi^{\rm AGN}_{\gamma}}{\sigma^{\rm{data}}}  \right]^2.
    \label{equ: chi2_all}
\ee
It is evident that the model incorporating the DPL background enforces stricter constraints in all scenarios compared to those without it. This suggests that a more profound comprehension of the astrophysical background sources could lead to a substantial enhancement in setting upper limits on the presence of PBHs. The capability to exclude a significant portion of the dataset with the DPL model results in only a small fraction of the flux being attributed to PBHs, thereby directly boosting the constraint ability. However, it is important to note that the DPL form is purely phenomenological, underscoring the necessity to gain a better understanding of the astrophysical background in future analyses.

It appears that by utilizing Equ.(\ref{equ: fraction}), we can ascertain the value of $P$, considering $f_{\rm Ps}$ as a variable parameter. Previous works \cite{Korwar:2023kpy, Iguaz:2021irx} have neglected the contribution from IA due to its minor impact, leading to a slight discrepancy in their expression of $f_{\rm Ps}$. Therefore, in this analysis, we set $f_{\rm Ps}$ as either 0 or 1 and leverage the fractions from Equ.(\ref{equ: fraction}) to weigh each emission source. The constraints on $f_{\rm Ps}$ are depicted in Fig.\ref{fig: bounds_fPs}, where $f_{\rm Ps} = 1$ corresponds to scenarios with no spin PBHs in the presence of a DPL background (solid blue line) and only PBH model (dashed blue line), while $f_{\rm Ps} = 0$ is applied to both models (with DPL is solid black line and without DPL is dashed black line).

It is observed that when the environmental temperature exceeds $8000$ K, positrons are more likely to undergo direct annihilation with both free and bound electrons rather than forming Ps. Therefore, the scenario with $f_{\rm Ps}=1$ represents a more realistic case under Milky Way conditions \cite{Guessoum:2005cb}. In Fig.\ref{fig: bounds_fPs}, the outcomes for $f_{\rm Ps}=0$ exhibit stronger constraints, indicating a potential underestimation of the contribution from the EG sources in the existing literature.

\subsection{Considering of data selection}
It is worth noting that, as discussed in ref.\cite{Korwar:2023kpy}, the error bars associated with the SMM measurement of the isotropic X-ray flux are considered to be unreliable. Consequently, we also present an analysis based solely on SMM data. The remaining dataset can be categorized into three scenarios: utilizing only SMM data, utilizing all data except SMM, and utilizing all data including SMM. The $\chi^2$ calculation follows the same formula as Equ.(\ref{equ: chi2_all}), with varying selections of data for $\phi_\gamma^{\rm data}$ and their corresponding uncertainties $\sigma_\gamma^{\rm data}$. The results are compared in Fig.\ref{fig: bounds_woSMM}. It is observed that SMM data in isolation can effectively constrain a significant portion of the considered PBH masses. Consequently, the outcomes from only using SMM data and employing all data with SMM align closely across most of the range. As depicted in Fig.\ref{fig: CXB}, SMM data holds particular relevance for higher energies surpassing approximately $200$ keV. However, for more massive PBHs, the spectral characteristics tend to soften, rendering the data without SMM more crucial. These findings underscore the importance of utilizing data within appropriate energy ranges with minimal error bars to yield robust constraints.

In Fig.\ref{fig: bounds_woSMM}, it is evident that for PBHs with masses exceeding $1\times 10^{18}$ g, not only does the energy peak decrease and fall below $100$ keV, but it also significantly weakens. This indicates that high-energy data sources like SMM may not effectively constrain this mass range, thereby posing a challenge in directly achieving this objective using current keV data.

\subsection{Considering of PBHs spin}
The concept of a rotated PBH potentially leading to the generation of additional particles during Hawking evaporation is intriguing. In principle, it is assumed that PBHs do not possess spins at the time of their formation. Following their formation, PBHs are expected to remain stable, with only a small fraction, roughly one in a million PBHs, being formed with spins $\gtrsim 0.8$ \cite{PhysRevD.104.083018}. The existence and implications of rotated PBHs remain a topic of debate, warranting further exploration to better understand their implications. To shed light on this, a comparable analysis is presented within the context of the monochromatic mass function scenario discussed earlier. By adjusting the intrinsic spin towards the maximum value of $a=0.9999$, which corresponds to the maximum spin value in \texttt{BlackHawk}, the results pertaining to rotating PBHs exhibit significant enhancements. This is attributed to the higher number density resulting from the evaporation process in this particular scenario.

\begin{figure}[htp]
    \centering
    \includegraphics[width=9cm]{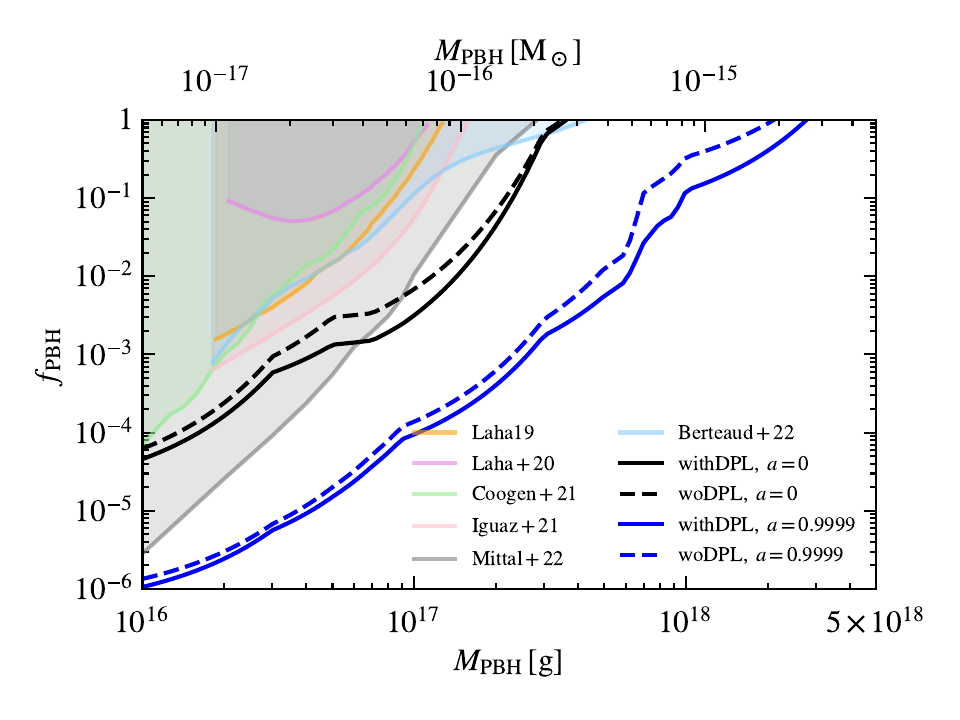}
    \caption{\label{fig: bounds_wobkg}Bounds derived on the PBHs abundance as a function of PBH mass, considering a monochromatic mass distribution and $f_{\rm Ps}=1$ are illustrated through four distinct cases: No spin scenario with a DPL background is represented by the black solid line; No spin scenario without a DPL background is depicted by the black dashed line; spin factor $a=0.9999$ with a DPL background is shown in the blue solid line, and spin factor $a=0.9999$ without a DPL background is displayed as the blue dashed line. These cases provide a comprehensive overview of the PBH abundance constraints under different assumptions regarding spin factors and the presence of a DPL background.}     
\end{figure}

It is fascinating to note the enhancements in constraints depicted in Fig.\ref{fig: bounds_wobkg} when considering the influence of PBH rotation. These results indicate an expansion in the covered parameter space, leading to improved constraints. Notably, even in the most conservative scenarios where there is no spin and no background modeling, we observe better limits, especially in the $10^{17}$ g range. This improvement can be attributed to the comprehensive consideration of all potential emissions in our calculations, resulting in a slight increase in the total flux and subsequently refining the bounds on PBH abundance.

\section{\label{sec: concl}Conclusion}
The article delves into the investigation of whether PBHs within the mass range of $10^{16}$ to $5\times 10^{18}$ g, characterized by a monochromatic mass distribution, could potentially account for all or part of dark matter. To establish the most stringent constraint line, various potential photon sources were considered, encompassing emissions in the X-ray and soft $\gamma$-ray spectra, as well as those arising from electron or positron annihilation radiation processes involving QED interactions leading to electron-positronium formation and subsequent photon emissions.

By differentiating between extragalactic and Galactic contributions and assigning fractions to each process comprehensively, a more detailed analysis was conducted. The study revealed that the total flux generated can exceed the individual components typically addressed in previous literature. The data were evaluated considering solely the contribution from PBHs and a more realistic scenario incorporating potential astrophysical background models. Consequently, the most conservative constraints were found to be more robust compared to previous bounds.
The findings suggest the exclusion of the entirety of DM being comprised of PBHs within the mass range of $2.5\times 10^{17}-3 \times 10^{17}$ g when accounting for the DPL background. PBHs exhibiting extreme spins exhibit limited potential to constitute DM.

Observations in the X-ray and soft $\gamma$-ray bands have the capability to probe PBHs down to the size of asteroids. Furthermore, advancements in resolving astrophysical backgrounds and upcoming precise diffuse data hold promise for imposing stronger constraints on the yet unexplored window of PBHs. We also aim to integrate multi-wavelength data in an upcoming paper, to shrink and even close the gap in the PBH parameter space.

\section*{acknowledgments}
We are appreciated the help from Roberto Gilli for the CXB data. J.-Q.X. is supported by the National Science Foundation of China under grant Nos. U1931202 and 12021003, the National Key R\&D Program of China No. 2020YFC2201603, and the Fundamental Research Funds for the Central Universities.


\begin{thebibliography}{}
\expandafter\ifx\csname natexlab\endcsname\relax\def\natexlab#1{#1}\fi
\providecommand{\url}[1]{\href{#1}{#1}}
\providecommand{\dodoi}[1]{doi:~\href{http://doi.org/#1}{\nolinkurl{#1}}}
\providecommand{\doeprint}[1]{\href{http://ascl.net/#1}{\nolinkurl{http://ascl.net/#1}}}
\providecommand{\doarXiv}[1]{\href{https://arxiv.org/abs/#1}{\nolinkurl{https://arxiv.org/abs/#1}}}

\bibitem[{Ajello {et~al.}(2008)Ajello, Greiner, Sato, Willis, Kanbach, Strong, Diehl, Hasinger, Gehrels, Markwardt, \& Tueller}]{Ajello:2008xb}
Ajello, M., Greiner, J., Sato, G., {et~al.} 2008, The Astrophysical Journal, 689, 666, \dodoi{10.1086/592595}

\bibitem[{Arbey \& Auffinger(2019)}]{Arbey:2019mbc}
Arbey, A., \& Auffinger, J. 2019, Eur. Phys. J. C, 79, 693, \dodoi{10.1140/epjc/s10052-019-7161-1}

\bibitem[{Auffinger \& Arbey(2021)}]{Auffinger:2020ztk}
Auffinger, J., \& Arbey, A. 2021, PoS, TOOLS2020, 024, \dodoi{10.22323/1.392.0024}

\bibitem[{Ballesteros {et~al.}(2020)Ballesteros, Coronado-Bl\'azquez, \& Gaggero}]{Ballesteros:2019exr}
Ballesteros, G., Coronado-Bl\'azquez, J., \& Gaggero, D. 2020, Phys. Lett. B, 808, 135624, \dodoi{10.1016/j.physletb.2020.135624}

\bibitem[{Bambi {et~al.}(2008)Bambi, Dolgov, \& Petrov}]{Bambi:2008kx}
Bambi, C., Dolgov, A.~D., \& Petrov, A.~A. 2008, Phys. Lett. B, 670, 174, \dodoi{10.1016/j.physletb.2009.10.053}

\bibitem[{Beacom {et~al.}(2005)Beacom, Bell, \& Bertone}]{beacom2005gamma}
Beacom, J.~F., Bell, N.~F., \& Bertone, G. 2005, Physical Review Letters, 94, 171301

\bibitem[{Beacom \& Yuksel(2006)}]{Beacom:2005qv}
Beacom, J.~F., \& Yuksel, H. 2006, Physical Review Letters, 97, 071102, \dodoi{10.1103/PhysRevLett.97.071102}

\bibitem[{Bringmann {et~al.}(2012)Bringmann, Huang, Ibarra, Vogl, \& Weniger}]{Bringmann:2012vr}
Bringmann, T., Huang, X., Ibarra, A., Vogl, S., \& Weniger, C. 2012, Journal of Cosmology and Astroparticle Physics, 07, 054, \dodoi{10.1088/1475-7516/2012/07/054}

\bibitem[{Cai {et~al.}(2021)Cai, Ding, Yang, \& Zhou}]{cai_constraints_2021}
Cai, R.-G., Ding, Y.-C., Yang, X.-Y., \& Zhou, Y.-F. 2021, Journal of Cosmology and Astroparticle Physics, 03, 057, \dodoi{10.1088/1475-7516/2021/03/057}

\bibitem[{Carr {et~al.}(2016)Carr, K{\"u}hnel, \& Sandstad}]{carr_primordial_2016}
Carr, B., K{\"u}hnel, F., \& Sandstad, M. 2016, Physical Review D, 94, 083504, \dodoi{10.1103/PhysRevD.94.083504}

\bibitem[{Carr {et~al.}(2017)Carr, Raidal, Tenkanen, Vaskonen, \& Veerm\"ae}]{Carr:2017jsz}
Carr, B., Raidal, M., Tenkanen, T., Vaskonen, V., \& Veerm\"ae, H. 2017, Phys. Rev. D, 96, 023514, \dodoi{10.1103/PhysRevD.96.023514}

\bibitem[{Chen {et~al.}(2022)Chen, Zhang, \& Long}]{Chen:2021ngo}
Chen, S., Zhang, H.-H., \& Long, G. 2022, Physical Review D, 105, 063008, \dodoi{10.1103/PhysRevD.105.063008}

\bibitem[{Chongchitnan \& Silk(2021)}]{PhysRevD.104.083018}
Chongchitnan, S., \& Silk, J. 2021, Phys. Rev. D, 104, 083018, \dodoi{10.1103/PhysRevD.104.083018}

\bibitem[{Churazov {et~al.}(2007)Churazov, Sunyaev, Revnivtsev, Sazonov, Molkov, Grebenev, Winkler, Parmar, Bazzano, Falanga, Gros, Lebrun, Natalucci, Ubertini, Roques, Bouchet, Jourdain, Kn{\"o}dlseder, Diehl, {Budtz-Jorgensen}, Brandt, Lund, Westergaard, Neronov, T{\"u}rler, Chernyakova, Walter, Produit, Mowlavi, {Mas-Hesse}, Domingo, Gehrels, Kuulkers, Kretschmar, \& Schmidt}]{Churazov:2006bk}
Churazov, E., Sunyaev, R., Revnivtsev, M., {et~al.} 2007, Astronomy \& Astrophysics, 467, 529, \dodoi{10.1051/0004-6361:20066230}

\bibitem[{Coogan {et~al.}(2021)Coogan, Morrison, \& Profumo}]{Coogan:2020tuf}
Coogan, A., Morrison, L., \& Profumo, S. 2021, Physical Review Letters, 126, 171101, \dodoi{10.1103/PhysRevLett.126.171101}

\bibitem[{DeRocco \& Graham(2019)}]{DeRocco:2019fjq}
DeRocco, W., \& Graham, P.~W. 2019, Phys. Rev. Lett., 123, 251102, \dodoi{10.1103/PhysRevLett.123.251102}

\bibitem[{Franceschini {et~al.}(2008)Franceschini, Rodighiero, \& Vaccari}]{Franceschini:2008tp}
Franceschini, A., Rodighiero, G., \& Vaccari, M. 2008, Astron. Astrophys., 487, 837, \dodoi{10.1051/0004-6361:200809691}

\bibitem[{Fukada {et~al.}(1975)Fukada, Hayakawa, Ikeda, Kasahara, Makino, \& Tanaka}]{fukada_rocket_1975}
Fukada, Y., Hayakawa, S., Ikeda, M., {et~al.} 1975, Astrophysics and Space Science, 32, L1, \dodoi{10.1007/BF00646232}

\bibitem[{{Gendreau} {et~al.}(1995){Gendreau}, {Mushotzky}, {Fabian}, {Holt}, {Kii}, {Serlemitsos}, {Ogasaka}, {Tanaka}, {Bautz}, {Fukazawa}, {Ishisaki}, {Kohmura}, {Makishima}, {Tashiro}, {Tsusaka}, {Kunieda}, {Ricker}, \& {Vanderspek}}]{Gendreau95}
{Gendreau}, K.~C., {Mushotzky}, R., {Fabian}, A.~C., {et~al.} 1995, Publications of the Astronomical Society of Japan, 47, L5

\bibitem[{{Georgantopoulos} {et~al.}(1991){Georgantopoulos}, {Stewart}, {Pounds}, {Shanks}, {Boyle}, \& {Griffiths}}]{Georgantopoulos91}
{Georgantopoulos}, I., {Stewart}, G.~C., {Pounds}, K.~A., {et~al.} 1991, in Astronomical Society of the Pacific Conference Series, Vol.~21, The Space Distribution of Quasars, ed. D.~{Crampton}, 6

\bibitem[{Gruber {et~al.}(1999)Gruber, Matteson, Peterson, \& Jung}]{Gruber:1999yr}
Gruber, D.~E., Matteson, J.~L., Peterson, L.~E., \& Jung, G.~V. 1999, The Astrophysical Journal, 520, 124, \dodoi{10.1086/307450}

\bibitem[{Guessoum {et~al.}(2005)Guessoum, Jean, \& Gillard}]{Guessoum:2005cb}
Guessoum, N., Jean, P., \& Gillard, W. 2005, Astronomy \& Astrophysics, 436, 171, \dodoi{10.1051/0004-6361:20042454}

\bibitem[{Hawking(1971)}]{10.1093/mnras/152.1.75}
Hawking, S. 1971, Monthly Notices of the Royal Astronomical Society, 152, 75, \dodoi{10.1093/mnras/152.1.75}

\bibitem[{{Hawking}(1975)}]{1975CMaPh..43..199H}
{Hawking}, S.~W. 1975, Communications in Mathematical Physics, 43, 199, \dodoi{10.1007/BF02345020}

\bibitem[{Iguaz {et~al.}(2021)Iguaz, Serpico, \& Siegert}]{Iguaz:2021irx}
Iguaz, J., Serpico, P.~D., \& Siegert, T. 2021, Physical Review D, 103, 103025, \dodoi{10.1103/PhysRevD.103.103025}

\bibitem[{Keith \& Hooper(2021)}]{Keith:2021guq}
Keith, C., \& Hooper, D. 2021, Phys. Rev. D, 104, 063033, \dodoi{10.1103/PhysRevD.104.063033}

\bibitem[{Keith {et~al.}(2022)Keith, Hooper, Linden, \& Liu}]{Keith:2022sow}
Keith, C., Hooper, D., Linden, T., \& Liu, R. 2022, Physical Review D, 106, 043003, \dodoi{10.1103/PhysRevD.106.043003}

\bibitem[{Korwar \& Profumo(2023)}]{Korwar:2023kpy}
Korwar, M., \& Profumo, S. 2023, Journal of Cosmology and Astroparticle Physics, 2023, 054, \dodoi{10.1088/1475-7516/2023/05/054}

\bibitem[{Kushino {et~al.}(2002)Kushino, Ishisaki, Morita, Yamasaki, Ishida, Ohashi, \& Ueda}]{Kushino:2002vk}
Kushino, A., Ishisaki, Y., Morita, U., {et~al.} 2002, Publications of the Astronomical Society of Japan, 54, 327, \dodoi{10.1093/pasj/54.3.327}

\bibitem[{Laha(2019)}]{Laha:2019ssq}
Laha, R. 2019, Phys. Rev. Lett., 123, 251101, \dodoi{10.1103/PhysRevLett.123.251101}

\bibitem[{Laha {et~al.}(2020)Laha, Mu{\~n}oz, \& Slatyer}]{laha_integral_2020}
Laha, R., Mu{\~n}oz, J.~B., \& Slatyer, T.~R. 2020, Physical Review D, 101, 123514, \dodoi{10.1103/PhysRevD.101.123514}

\bibitem[{Manohar \& Ruiz-Femenia(2004)}]{Manohar:2003xv}
Manohar, A.~V., \& Ruiz-Femenia, P. 2004, Phys. Rev. D, 69, 053003, \dodoi{10.1103/PhysRevD.69.053003}

\bibitem[{Navarro {et~al.}(1996)Navarro, Frenk, \& White}]{Navarro:1995iw}
Navarro, J.~F., Frenk, C.~S., \& White, S. D.~M. 1996, Astrophys. J., 462, 563, \dodoi{10.1086/177173}

\bibitem[{Ore \& Powell(1949)}]{Ore:1949te}
Ore, A., \& Powell, J.~L. 1949, Physical Review, 75, 1696, \dodoi{10.1103/PhysRev.75.1696}

\bibitem[{Prantzos {et~al.}(2011)Prantzos, Boehm, Bykov, Diehl, Ferriere, Guessoum, Jean, Knoedlseder, Marcowith, Moskalenko, Strong, \& Weidenspointner}]{Prantzos:2010wi}
Prantzos, N., Boehm, C., Bykov, A.~M., {et~al.} 2011, Reviews of Modern Physics, 83, 1001, \dodoi{10.1103/RevModPhys.83.1001}

\bibitem[{Rasera {et~al.}(2006)Rasera, Teyssier, Sizun, Cordier, Paul, Casse, \& Fayet}]{Rasera:2005sa}
Rasera, Y., Teyssier, R., Sizun, P., {et~al.} 2006, Phys. Rev. D, 73, 103518, \dodoi{10.1103/PhysRevD.73.103518}

\bibitem[{Ray {et~al.}(2021)Ray, Laha, Mu{\~n}oz, \& Caputo}]{ray_near_2021}
Ray, A., Laha, R., Mu{\~n}oz, J.~B., \& Caputo, R. 2021, Physical Review D, 104, 023516, \dodoi{10.1103/PhysRevD.104.023516}

\bibitem[{Sheth \& Tormen(1999)}]{Sheth:1999mn}
Sheth, R.~K., \& Tormen, G. 1999, Mon. Not. Roy. Astron. Soc., 308, 119, \dodoi{10.1046/j.1365-8711.1999.02692.x}

\bibitem[{Siegert {et~al.}(2022)Siegert, Boehm, Calore, Diehl, Krause, Serpico, \& Vincent}]{Siegert:2021upf}
Siegert, T., Boehm, C., Calore, F., {et~al.} 2022, Mon. Not. Roy. Astron. Soc., 511, 914, \dodoi{10.1093/mnras/stac008}

\bibitem[{Tan {et~al.}(2022)Tan, Yan, Qiu, \& Xia}]{Tan:2022lbm}
Tan, X.-H., Yan, Y.-J., Qiu, T., \& Xia, J.-Q. 2022, Astrophys. J. Lett., 939, L15, \dodoi{10.3847/2041-8213/ac9668}

\bibitem[{Ueda {et~al.}(2014)Ueda, Akiyama, Hasinger, Miyaji, \& Watson}]{Ueda:2014tma}
Ueda, Y., Akiyama, M., Hasinger, G., Miyaji, T., \& Watson, M.~G. 2014, The Astrophysical Journal, 786, 104, \dodoi{10.1088/0004-637X/786/2/104}

\bibitem[{Vecchi {et~al.}(1999)Vecchi, Molendi, Guainazzi, Fiore, \& Parmar}]{Vecchi:1999ph}
Vecchi, A., Molendi, S., Guainazzi, M., Fiore, F., \& Parmar, A.~N. 1999, Astron. Astrophys., 349, L73.
\newblock \doarXiv{astro-ph/9908323}

\bibitem[{{Watanabe} {et~al.}(1997){Watanabe}, {Hartmann}, {Leising}, {The}, {Share}, \& {Kinzer}}]{Watanabe97}
{Watanabe}, K., {Hartmann}, D.~H., {Leising}, M.~D., {et~al.} 1997, in American Institute of Physics Conference Series, Vol. 410, Proceedings of the Fourth Compton Symposium, ed. C.~D. {Dermer}, M.~S. {Strickman}, \& J.~D. {Kurfess}, 1223--1227, \dodoi{10.1063/1.53933}

\bibitem[{Weidenspointner {et~al.}(2000)}]{Weidenspointner:2000aq}
Weidenspointner, G., {et~al.} 2000, AIP Conf. Proc., 510, 581, \dodoi{10.1063/1.1303269}

\end{thebibliography}

\end{document}